\documentclass[11pt]{article}

\usepackage{cite}
\usepackage[pdftex]{graphicx}
\usepackage[cmex10]{amsmath}
\usepackage{array}
\usepackage[textfont=small]{caption}
\usepackage[textfont=small]{subfig}%\usepackage[tight,footnotesize]{subfigure}
\usepackage{url}

\usepackage{amssymb}
\usepackage{amsthm}
\usepackage{verbatim}

\usepackage[left=1.25in,right=1.25in,top=1.25in,bottom=1in]{geometry}%,nofoot,nohead

\newcommand{\appalone} %appendix as a standalone doc instead of embedded

\begin{document}

\title{Achieving Energy Efficiency for Altruistic DISH: Three Properties}
\author{Tony T. Luo\smallskip\\National University of Singapore}
\date{\vspace{-4ex}}

\maketitle
%\usepackage{fancyhdr}
%\pagestyle{fancy}
%\thispagestyle{fancy}
%\lhead{IEEE TRANSACTIONS ON MOBILE COMPUTING, April 2012}
%\renewcommand{\headrulewidth}{0pt}

\newcommand{\fref}[1]{Fig.~\ref{#1}}
\newcommand{\sref}[1]{Section~\ref{#1}}
\newcommand{\aref}[1]{appendix~\ref{#1}}
\newcommand{\tref}[1]{Table~\ref{#1}}
\newcommand{\pref}[1]{Prop.~\ref{#1}}
\newcommand{\mref}[1]{Theorem~\ref{#1}}
\newcommand{\dref}[1]{Def.~\ref{#1}}
\newtheorem{thm}{Theorem}
\newtheorem{defn}{Definition}
\newtheorem{prop}{Proposition}

\begin{abstract}
In an altruistic DISH protocol, additional nodes called ``altruists'' are deployed in a multi-channel ad hoc network to achieve energy efficiency while still maintaining the original throughput-delay performance. The responsibility of altruists is to constantly monitor the control channel and awaken other (normal) nodes when necessary (to perform data transmissions). Altruists never sleep while other nodes sleep as far as possible. This technical report proves three properties related to this cooperative protocol. The first is the conditions for forming an unsafe pair (UP) in an undirected graph. The second is the necessary and sufficient conditions for full cooperation coverage to achieve the void of multi-channel coordination (MCC) problems. The last is the NP-hardness of determining the minimum number and locations of altruistic nodes to achieve full cooperation coverage.
\end{abstract}

\section{Introduction}

The main background is a DISH-based protocol called DISH-p. Here we describe this protocol according to \cite{luo12tmc-altruist}. In DISH-p, a sender and a receiver set up communication using PRA/PRB packets and then confirm using CFA/CFB packets. A neighbor will send INV packet if it identifies an MCC problem via the information conveyed by PRA/PRB.

\begin{figure}[!ht]
\centering
\subfloat[Control channel handshake.]
{\includegraphics[width=.7\linewidth]{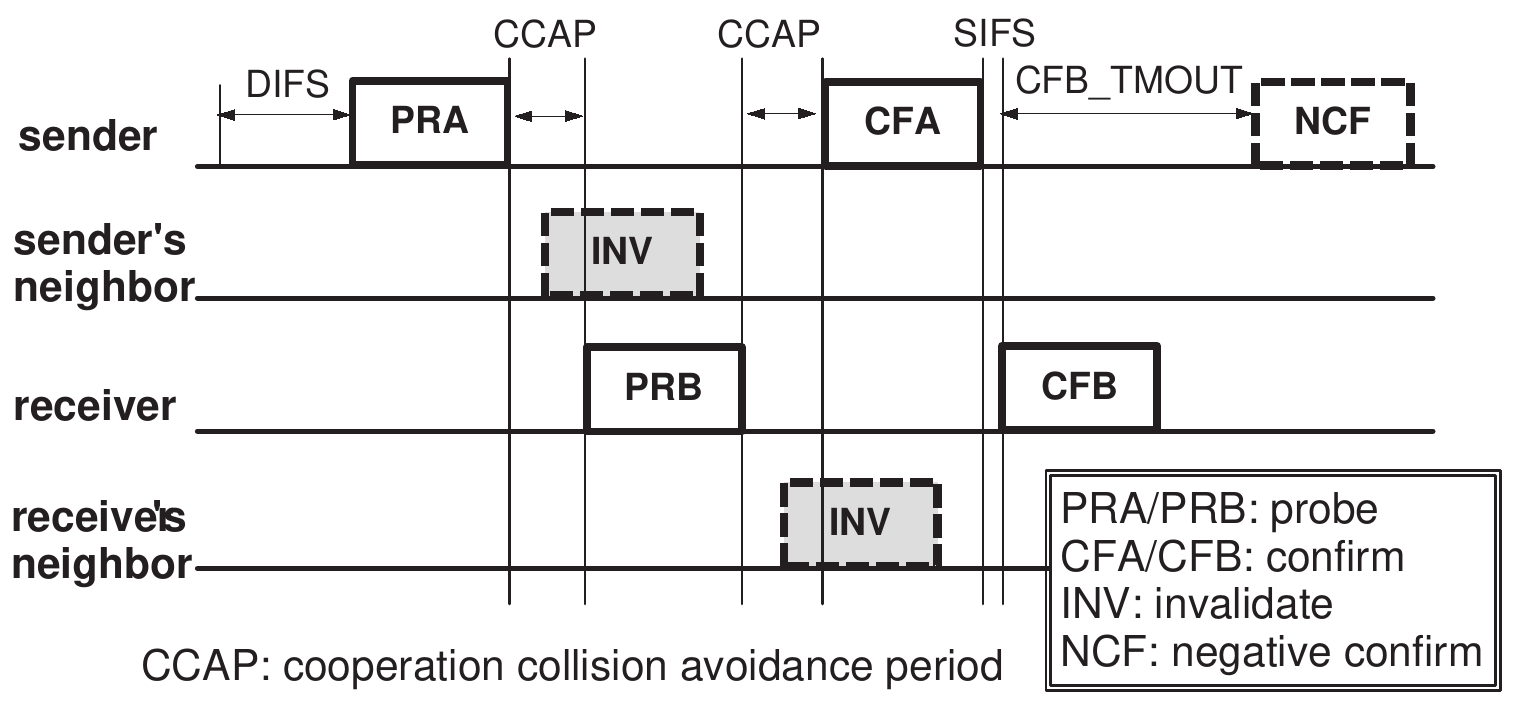}\label{fig:hsk-ctrl}}\vfil
\subfloat[Frame format. INV carries the channel usage information of an established and ongoing data exchange on a data channel (which engages the ``deaf'' receiver in the case of deaf terminal problem).]
{\makebox[\linewidth]{\includegraphics[width=.5\linewidth]{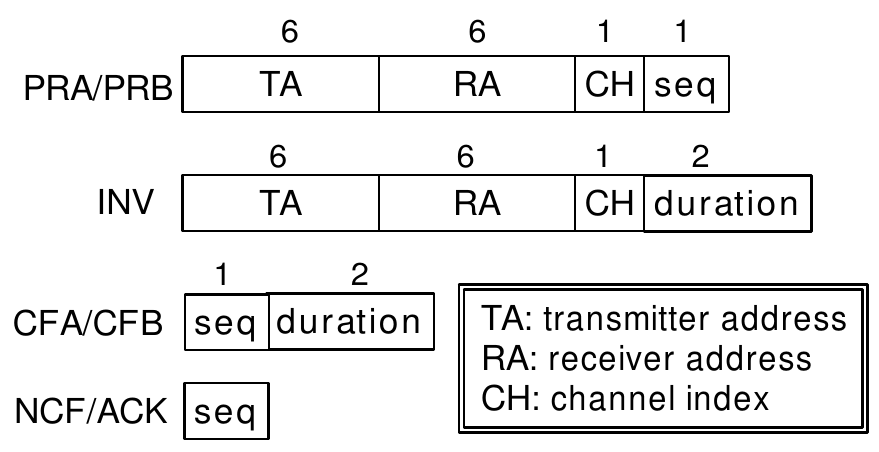}\label{fig:format}}}\vfil
\subfloat[Channel usage table. Each node maintains one to cache its overheard control information.]
{\makebox[\linewidth]{\includegraphics[width=.3\linewidth]{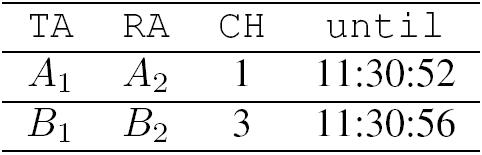}\label{fig:chtab}}}
\caption{Elements of the DISH-p protocol.}\label{fig:dishp}
\end{figure}

There is one control channel and multiple data channels. On the control channel, a sender and a receiver exchange PRA/PRB (see \fref{fig:hsk-ctrl}) to select a data channel, and then exchange CFA/CFB to confirm the channel selection. The frame format is shown in \fref{fig:format}. If a neighbor identifies an MCC problem (via PRA or PRB), it will prepare to send an INV packet, during a cooperation collision avoidance period (CCAP), to alarm the sender or the receiver to back off. If there is no MCC problem identified by any neighbor (no INV will be sent), the sender and the receiver will switch to their chosen data channel and start DATA/ACK exchange. During DIFS and CCAP, carrier sensing is turned on to mitigate collisions via CSMA.

%Figure pushed to the end

CCAP is introduced to mitigate the collision of multiple simultaneously sent INVs. A neighbor who identifies an MCC problem wil send INV only if it senses the control channel to be free for a period of \textsl{Uniform}[0, CCAP]. Hence a neighbor who sends INV will suppress its neighbors via CSMA.\footnote{CSMA does not avoid all collisions because not all the neighbors may hear each other. However, a collision of such still conveys an alarm to the sender/receiver because INV represents a {\em negative} message, and hence the sender/receiver will still back off. What is only compromised is that the sender/receiver will not know precisely how long at least it should back off and hence will have to estimate a backoff period, which is not a serious problem.} NCF is sent when the sender waits for CFB until timeout (due to the receiver receiving INV), in order to inform the sender's neighbors to disregard CFA.

The applicable scenarios of the protocol are mesh networks and ad hoc networks, not sensor networks. In sensor networks, data packets are usually small and the overhead of the control channel handshake will be significant. Even using a packet train would not suit because sensing traffic is usually periodic and not bursty.

\subsection{Altruistic DISH}

The above protocol was originally CAM-MAC introduced in \cite{cammac09tmc}. In the work
\cite{luo12tmc-altruist}, two energy-efficient strategies was investigated and this technical report focuses on the advocated strategy called altruistic DISH. In this strategy, additional nodes called {\em altruists} are deployed to take over the responsibility of information sharing (i.e., cooperation) from the existing nodes, which we call {\em peers} to distinguish from altruists, so that peers can sleep when idle. Altruists are the same as peers in terms of hardware, but are different in terms of software: they solely cooperate (do not carry data traffic) and always stay awake.

An apparent drawback of this strategy is that it requires additional nodes. However, this is offset by substantive advantages. First, it is very simple to implement the strategy: one only needs to introduce a boolean flag to disable data related functions on altruists and cooperation related functions on peers. We have done this in both our simulation code and hardware implementation code. Equally importantly, there is no additional runtime mechanism and hence runtime overhead.

Second, unlike the in-situ strategy, this strategy does not have the multi-channel broadcasting problem. Altruists always stay on the same channel (control channel) and send/receive packets only on the control channel.

Third, this strategy is robust to network dynamics (such as traffic and residual energy). Every altruist is cooperative and will react to every MCC problem that it identifies; they do not need to adjust any parameter on the fly. In fact, even the deployment of altruists, which is an offline process, can be done with a constant number for any given peer density.

Fourth, since peers only carry data traffic and need not to cooperate, they are like nodes in traditional (non-DISH) networks and thus can adopt a legacy sleep-wake scheduling algorithm, where a lot of choices are available in prior work.

Finally, unlike the in-situ strategy and the original DISH where cooperation is provided in an {\em opportunistic} manner---meaning that cooperative nodes are not always available, altruistic DISH provides cooperation in a {\em guaranteed} manner.

\section{Three Properties for Altruistic DISH}

\subsection{Forming unsafe pairs}
%{\bf Proposition: }
\begin{prop}\label{prop:form_up}
In an undirected graph where each vertex represents a peer and each edge represents the relationship between two neighboring peers, denote by $d_i$ the degree of an arbitrary vertex $i$. If PSM is not used, two adjacent vertices $i$ and $j$ form an UP if and only if:
\begin{enumerate}
\renewcommand{\labelenumi}{(\alph{enumi})}
\item $d_i\ge 2$, $d_j\ge 2$, and $d_i=d_j=2$ does not hold, or
\item $d_i=d_j=2$, and $i$ and $j$ are not on the same three-cycle (i.e., triangle).
\end{enumerate}
If PSM is used (peers sleep when idle), the above condition remains unchanged for the channel conflict problem, but changes to the following for the deaf terminal problem:

$d_i\ge 1$, $d_j\ge 1$, and $d_i=d_j=1$ does not hold.
\end{prop}

The above quotes Proposition 1 from \cite{luo12tmc-altruist}.

\begin{proof}
First consider the case without PSM.

{\it Sufficiency}: If condition (a) or (b) is satisfied, $i$ and $j$ form two independent communicable pairs, say $p_i$ ($i,i'$) and $p_j$ ($j,j'$), as illustrated in \fref{fig:form_up}. Suppose $p_i$ switches to a data channel $ch_i$ when $p_j$ is communicating on data channel $ch_j$, then this channel usage of $ch_i$ is unknown to $j$. After $p_j$ switches back to the control channel and if $j$ initiates another communication while $p_i$ is still communicating on $ch_i$, then (i) a channel conflict problem is created if $j$ choose to use channel $ch_i$, or (ii) a deaf terminal problem is created if $j$ initiates this communication with $i$.

\begin{figure}[t]
\centering
\includegraphics[trim=0 2mm 0 3mm,clip,width=0.8\linewidth]{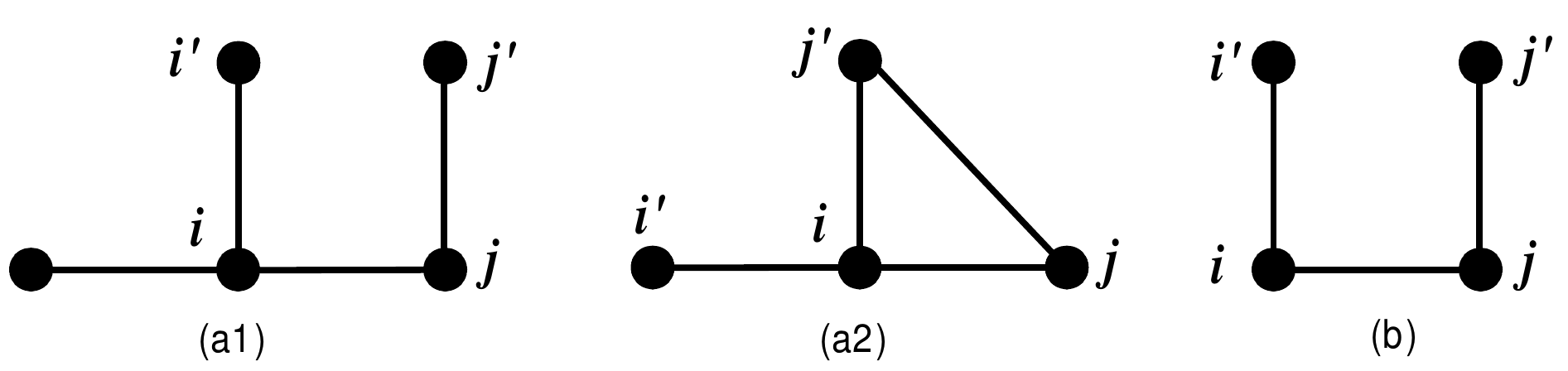}
\caption{Edges represent neighboring relationships. Corresponding to \ifdefined\appalone Proposition 1, \else \pref{prop:form_up}, \fi subfigures (a1) and (a2) illustrate condition (a), subfigure (b) illustrates condition (b).}
\label{fig:form_up}
\end{figure}
%(the caption if leaving out proof:) Figure (a1) and (a2) correspond to condition (a), where nodes $i$ and $j$ can form two respective communication pairs: ($i,u/w$) and ($j,v$) in (a1), ($i,w$) and ($j,u$) in (a2). Figure (b) corresponds to condition (b), where the two communication pairs are ($i,u$) and ($j,v$).

{\it Necessity}: Equivalently, we prove that if neither of the conditions (a) and (b) is satisfied, i.e., $d_i$ (or $d_j$) is 1 or $i$ and $j$ are on the same three-cycle, $i$ and $j$ does {\em not} form an UP. Since there is only one communicable pair, channel conflict problems will not be possible. Furthermore, whenever the (one) communicable pair performs a control channel handshake, the third node (if there is) will always be informed since we have assumed a node will always listen to the control channel when idle (if without PSM). Therefore, deaf terminal problems are also not possible.

Clearly, in the case with PSM, since there are at least three nodes, a sleeping node, say $i$, will miss the communication between $j$ and a third node. Hence a deaf terminal problem will be created when $i$ wakes up and initiates communication with $j$. Note that in this paper, deaf terminals are defined w.r.t. multi-channel; a sleeping receiver is not called a deaf terminal.
\end{proof}

\subsection{MCC-free condition}
\begin{prop}\label{prop:coll_free}
Consider a network using altruistic DISH. In order to achieve free of MCC problems, full cooperation coverage is
\begin{enumerate}
\item necessary for a multi-hop network, and
\item necessary and sufficient for a single-hop network.
\end{enumerate}
\end{prop}

The above quotes Theorem 2 from \cite{luo12tmc-altruist}.

\begin{proof}
{\it Necessity}: Since it is always possible for an UP to create MCC problems, an UP has to become a CUP to avoid these problems in a network using altruistic DISH. In other words, full cooperation coverage is a necessary condition for the network, irrespective of single-hop or multi-hop, to be free of MCC problems.

{\it Sufficiency}: In a single-hop network, one altruist achieves full cooperation coverage. Due to CSMA, each time only one control channel handshake can be accomplished. Therefore, every MCC problem created by such handshakes will be identified and prevented by the altruist. In case that there are more than one altruist, there is a marginal chance of collision between cooperative messages (in fact the chances are very low because of CCAP). However the proposition still holds because such collision still indicates an MCC problem as explained in \ifdefined\appalone Section II-A. \else \sref{sec:dish-detail}. \fi

{\bf Remark}: Full cooperation coverage is not a sufficient condition for multi-hop networks to be free of MCC problems, because concurrent and geographically distributed transmissions may overlap at altruists and hence not all MCC problems may be identified.
%%In addition, the reactive cooperation disallows any node to intervene a handshake without having identified a problem. Therefore, collision cannot be eliminated even if all peers are covered.
\end{proof}

\subsection{NP-hardness of altruistic nodes placement}
\begin{thm}\label{thm:nphard}
Consider a network with a given topology formed by peers on a finite plane. The problem of determining the minimum number and the locations of altruists to achieve full cooperation coverage, is NP-hard.
\end{thm}

The above quotes Proposition 2 from \cite{luo12tmc-altruist}.

\begin{proof}

{\bf Step 1: Identify UPs}

This step is to obtain a set $U$ of all the UPs in the network by identifying UPs according to \ifdefined\appalone Proposition 1. \else \pref{prop:form_up}. \fi As an example, see a six-node network shown in \fref{fig:faces} and we only consider the case without PSM for conciseness. There are three UPs, and $U=\{ (i,j), (j,k), (i,k) \}$.%\footnote{When using PSM {\em and} considering deaf terminal problems, $(i',i), (j',j), (k',k)$ are also UPs. We leave out this special case for a concise and more illustrative example.}

%\begin{wrapfig}{r}{0.36\linewidth}
\begin{figure}
\ifdefined\thesis
\centering\includegraphics[width=0.3\linewidth]{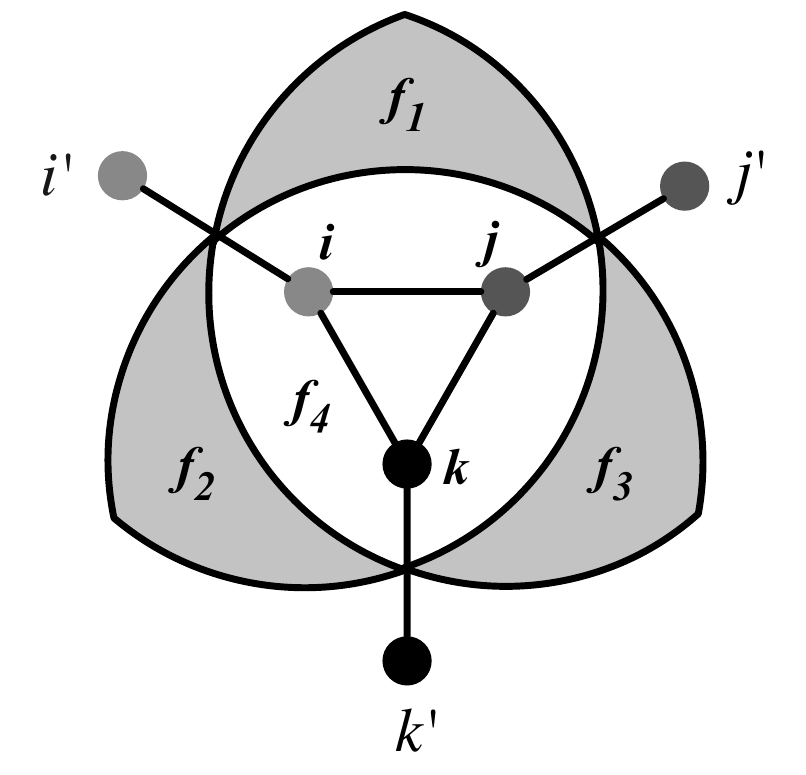}
\else
\centering\includegraphics[width=0.38\linewidth]{gray}
\fi
\caption{An illustration of \ifdefined\appalone Theorem 2. \else \mref{thm:nphard}. \fi Edges represent neighboring relationships, and arcs represent radio ranges of $i$, $j$ and $k$.}
\label{fig:faces}
\vspace{-4mm}\end{figure}

%\vspace{2mm}
{\bf Step 2: Construct Orphanage Set}

This step is to construct $\mathcal H=\{H_i|i=1,2,...p\}$ which is a set of all the orphanages in a network. To define orphanage, we first define {\it face}.

\begin{defn}
A {\em face} is a region bounded by the (circular) radio boundaries of the peers who form UPs (there is no boundary inside a face). We say that {\em a face covers an UP}, if an altruist on any point of this face covers this UP.
\end{defn}

For example, in \fref{fig:faces}, $i$, $j$ and $k$ are all the peers that form UPs, $f_1,f_2,f_3$ and $f_4$ are all the faces, where, e.g., $f_1$ covers UP $(i,j)$. Note that $f_1\cup f_4$ is not a face.

%Denote by $I_{ab}$ the intersection region of $a$ and $b$'s radio ranges, where $(a,b)$ is an arbitrary UP. Clearly, an altruist in $I_{a_1b_1}$ will cover $(a_1,b_1)$, and an altruist in $I_{a_1b_1}\cap I_{a_2b_2}$ will cover both $(a_1,b_1)$ and $(a_2,b_2)$. Eventually, an altruist in a {\em face} will cover a set of UPs irrespective of any specific {\em point} it is located at.

\begin{defn}
An {\em orphanage} is the maximum set of UPs covered by a face. Rigorously, an orphanage $H$ is a set of UPs ($H\subseteq U$) covered by a face $f_H$, and $\forall u\in U\setminus H$, $u$ is not covered by $f_H$.
\end{defn}

For example, in \fref{fig:faces}, $H_1=\{(i,j)\}$ and $H_4=\{ (i,j), (j,k), (i,k)\}$ are two orphanages covered by faces $f_1$ and $f_4$, respectively. But $H_4'=\{ (i,j), (i,k)\}$ is not an orphanage. There are totally four orphanages in \fref{fig:faces}.%though covered by $f_4$, because $(j,k)\notin H_4'$ is also covered by $f_4$.

%For each UP $(i,j)$, let $R_{ij}$ denote the largest region such that an altruist located at any point in $R_{ij}$ can cover the UP $(i,j)$. For a subset $C \subseteq U$, define its face as $f(C) = \cap_{(i,j) \in C}R_{ij}$. Then, a subset $C$ is a maximal clique, iff $f(C) \not= \{\phi\}$ and $\forall C\cup u, u \in U \setminus C, f(C \cup u) = \{\phi\}$. We call the face of a maximal clique a maximal face.

%faces are by partitioning a plane (without overlap) and an UP cannot be covered by two faces simultaneously, therefore
By definition, there is a one-to-one mapping between each orphanage and its covering face. Thus, finding all the orphanages in a network is equivalent to finding all the faces that covers at least one UP. This problem is the same as the target coverage problem \cite{coverage05} in sensor networks, and is shown by \cite{berman04} that the number of such faces is bounded by $|U|(|U|-1)+2$ and these faces can be found in time $O(|U|^3)$ by simply finding all the intersecting points of the circles (e.g., there are six such points in \fref{fig:faces}).

%\vspace{2mm}
{\bf Step 3: Formulate Problem}

With $U$ and $\mathcal H$, two problems can be posed:
\begin{enumerate}
\item Decision problem: given $U$, $\mathcal H$ and an integer $k$, determine whether a subset $\mathcal C=\{ H_i|i=1,2,...q\} \subseteq \mathcal H$ exists such that $\bigcup_{i=1}^q H_i =U$ and $q\le k$.
\item Optimization problem: given $U$ and $\mathcal H$, minimize $k=|\mathcal C|$ over all possible $\mathcal C=\{ H_i|i=1,2,...q\} \subseteq \mathcal H$, subject to $\bigcup_{i=1}^q H_i =U$.
\end{enumerate}
Since each orphanage $H_i\in \mathcal H$ corresponds to a unique face containing an altruist, $q$ ($q\le p$) is the minimum number of altruists that achieve full cooperation coverage.

The above two problems are the variants of the {\em set cover problem} defined by Karp~\cite{karp72}; the decision problem is NP-complete and the optimization problem is NP-hard. %Our problem is the optimization problem.
\end{proof}

\section{Conclusion}
This technical report proves three properties related to a cooperative, altruistic DISH protocol. These properties are regarding the conditions for forming an unsafe pair, the conditions for full cooperation coverage to achieve MCC free, and the NP-hardness of altruistic node deployment. On the other hand, this report focuses on the analysis of the ``add-on'' strategy (altruistic DISH) for the original cooperative protocol (DISH-p or CAM-MAC), while the reader is referred to \cite{mobihoc08,tmc10metric} for the analysis of the original protocol.

\end{document}